\documentclass[aps,pre,a4paper,10pt,floatfix,twoside,twocolumn,showpacs,showkeys]{revtex4-1}

\usepackage{amsmath,amssymb,bm,times,graphicx}

\usepackage{hyperref}
\hypersetup{unicode=true, pdftitle={Sensitivity to noise and ergodicity of an assembly line of cellular automata that classifies density}, pdfauthor={J. Ricardo G. Mendonça}, pdfkeywords={Density classification task, probabilistic cellular automata, Gacs-Kurdyumov-Levin, robustness, ergodicity, noise}, pdfsubject={Statistical Mechanics, Cellular Automata, Complexity}, pdfnewwindow=true, pdftoolbar=true, pdfmenubar=true, pdffitwindow=false, pdfstartview={FitH}, colorlinks=true, linkcolor=blue, citecolor=blue, urlcolor=blue, extension=pdf}

\def\leq{\leqslant}
\def\geq{\geqslant}

\begin{document}

\title{Sensitivity to noise and ergodicity of an assembly line of cellular automata that classifies density}

\author{J. Ricardo G. Mendon\c{c}a}
\email[Email: ]{jricardo@usp.br}
\affiliation{Instituto de F\'{\i}sica, Universidade de S\~{a}o Paulo -- 05314-970 S\~{a}o Paulo, SP, Brazil}

\begin{abstract}
We investigate the sensitivity of the composite cellular automaton of H.~Fuk\'{s} [\href{http://dx.doi.org/10.1103/PhysRevE.55.R2081}{Phys. Rev. E {\bf 55}, R2081 (1997)}] to noise and assess the density classification performance of the resulting probabilistic cellular automaton (PCA) numerically. We conclude that the composite PCA performs the density classification task reliably only up to very small levels of noise. In particular, it cannot outperform the noisy Gacs-Kurdyumov-Levin automaton, an imperfect classifier, for any level of noise. While the original composite CA is nonergodic, analyses of relaxation times indicate that its noisy version is an ergodic automaton, with the relaxation times decaying algebraically over an extended range of parameters with an exponent very close (possibly equal) to the mean-field value.
\end{abstract}



\maketitle


\section{\label{intro}Introduction}

The density (or majority) classification task for one-dimensional two-state cellular automata (CA) is a well known problem in theoretical computer science and, more generally, in the theory of complex systems \cite{wolfram,toffoli,sarkar}. The task consists of classifying binary strings according to their density $\rho$ of ones using local rules only, and it is completed successfully if a correct verdict as to whether $\rho < 1/2$ or $\rho > 1/2$ is obtained in finite time, at most linear in the size $L$ of the input string.

The density classification task originated in a paper by G\'{a}cs, Kurdyumov, and Levin (GKL) that was primarily concerned with providing evidence for the existence of local rules stabilizing a given phase against noise despite the ergodicity of the model \cite{gkl,maes}. For locally interacting systems of automata, the density classification task is a nontrivial task, because the cells have to achieve a global consensus cooperating locally. Systems with a controlling unit or layer such as a computer with a memory controller or a hierarchical neural network can overcome this difficulty in $O(L)$ or even $O(1)$ time by peeking each cell and accumulating its contents on a separate register. For one-dimensional systems of autonomous and memoryless cells interacting locally this is not an option; emergence of collective behavior is required in these cases.

It has been argued that the density classification task cannot be achieved without misclassifications by a single locally interacting two-state cellular automaton in any dimension \cite{belew}. Indeed, under the requirement that all the cells of the automaton must converge to the same state as the majority state in the initial configuration, no automata conceived to date could achieve $100\%$ efficiency. (For a critique on this classification criterion see \cite{capcar}.) The GKL automaton achieves a good $81.6\%$ performance in a sort of standard test consisting of classifying something between $10^{4}$ and $10^{7}$ random initial configurations of an $L=149$ array close to the ``critical'' density $\rho=1/2$. Improvements by humans over the GKL rule could achieve $82.2\%$ of correct classifications, and late genetic and co-evolution programming techniques were able to upgrade the success rate to $86.0\%$ \cite{physd,das,andre,crutch,juille,genetic}.

The no-go results mentioned above, however, did not rule out more complicated CA. In 1997, H.~Fuk\'{s} introduced a combination of CA that solves the density classification problem perfectly, with $100\%$ efficiency \cite{fuks97}. The composite CA of Fuk\'{s} corresponds to an ``assembly line'' of two CA (rules 184 and 232 in Wolfram's enumeration scheme; cf.~Sec.~\ref{model}), with one CA (184) running for the first half of the dynamics and the other CA (232) running for the second half. The time $T$ needed to complete the task is bounded by $T \leq L-2$. Moreover, when $L$ is even the assembly line CA deals with the case $\rho=1/2$ properly, by converging the initial configuration to the final configuration ${\cdots}\,010101\,{\cdots}$. Generalizations to arbitrary rational densities and to $n$-ary CA followed suit \cite{chau}.

In this article we investigate the classification performance of Fuk\'{s}'s assembly line CA under the influence of noise, turning the CA into a probabilistic cellular automaton (PCA) in which the transitions have a small probability of erring the final state. The classification efficiency of the PCA is assessed under several levels of noise in each of its components and for different system sizes, including even ones, by means of numerical experiments. We also compare the performance of the PCA with that of the noisy GKL automaton and provide evidence for its ergodicity at any level of noise.

This article is organized as follows. In Sec.~\ref{model} we introduce the assembly line CA together with its PCA version and describe our numerical experiments. In Sec.~\ref{results} we present our results and discuss their implications for the reliability of the PCA, including a discussion of its ergodicity. Finally, in Sec.~\ref{summary} we summarize our findings, make a few remarks and set forth some perspectives for future developments.


\section{\label{model}The assembly line CA and its PCA version}

A two-state cellular automaton transition function $\Phi$ on a one-dimensional array of $L$ cells is a map that evolves a given configuration $\bm{\eta}=(\eta_{1}, \eta_{2}, \ldots, \eta_{L}) \in \{0,1\}^{L}$ in discrete time $t \in \mathbb{N}$ according to 
\begin{equation}
\bm{\eta}(t+1)= \Phi(\bm{\eta}(t))= (\Phi_{1}(\bm{\eta}(t)), \ldots, \Phi_{L}(\bm{\eta}(t))),
\end{equation}
with each $\Phi_{i}(\bm{\eta}(t)) = \Phi_{i}(\eta_{i-l_{i}}(t), \ldots,$ $\eta_{i}(t),$ $\ldots, \eta_{i+r_{i}}(t))$ a function of $\{0,1\}^{l_{i}+r_{i}+1} \to \{0,1\}$.

A convenient way to express the local rules $\Phi_{i}$ when they are all equal is by means of a rule table. For the assembly line of Fuk\'{s}, we have two rule tables, one for each part of the assembly line. These rule tables are given in Table~\ref{tab:line}. We refer to this composite rule as CA $\Phi^{184}_{232}$. The numbers $184$ and $232$ come from reading the pattern of final states in the second row of each rule table as a binary number, after S.~Wolfram \cite{wolfram}. Since $\Phi^{184}_{232}$ is a composite CA, we must specify which part runs when and for how long. By definition, in this assembly line the CA~184 part runs for the first $T_{184} = \lfloor (L-2)/2 \rfloor$ time steps and the CA~232 part runs for the last $T_{232} = \lfloor (L-1)/2 \rfloor$ time steps \cite{fuks97}. Under these running times, $\Phi^{184}_{232}$ performs the density classification task perfectly and independently of $L$.

\begin{table}
\caption{\label{tab:line}Rule tables for the composite assembly line CA $\Phi^{184}_{232}$. For each CA, the first row lists the initial neighborhood and the second row lists the state reached by the central bit of the initial neighborhood under the action of the CA rules.}
\medskip
\begin{tabular}{ccccccccc}
\hline\hline
CA~184: & 111 & 110 & 101 & 100 & 011 & 010 & 001 & 000 \\ \hline
   {}   &  1  &  0  &  1  &  1  &  1  &  0  &  0  &  0  \\ \hline \hline \\
\hline\hline
CA~232: & 111 & 110 & 101 & 100 & 011 & 010 & 001 & 000 \\ \hline 
   {}   &  1  &  1  &  1  &  0  &  1  &  0  &  0  &  0  \\ \hline \hline
\end{tabular}
\end{table}

We are interested in the noisy version of $\Phi^{184}_{232}$, which can be obtained by allowing errors to intervene in the transitions made by $\Phi^{184}_{232}$. We thus introduce error rates $\varepsilon'$ and $\varepsilon''$, $0 \leq \varepsilon', \varepsilon'' \leq~\frac{1}{2}$, respectively to the CA~184 and CA~232 parts of the dynamics, turning CA $\Phi^{184}_{232}$ into PCA $\Phi^{184}_{232}(\varepsilon', \varepsilon'')$. Under this PCA, each cell $\eta_{i}(t)$ takes the value $\eta_{i}(t+1)$ that it would take under the noiseless $\Phi^{184}_{232}$ dynamics with probability $1-\varepsilon$ or its complement $1-\eta_{i}(t+1)$ with probability $\varepsilon$, where $\varepsilon$ equals $\varepsilon'$ or $\varepsilon''$ depending on which CA (184 or 232) is ruling the dynamics at instant $t$. In what follows, when the context is clear we sometimes refer to an unprimed $\varepsilon$ to denote either $\varepsilon'$ or $\varepsilon''$. The rule tables for this PCA are given in Table~\ref{tab:pca}.

The performance of a CA or PCA in the density classification task can be accounted as the fraction of initial configurations $\bm{\eta}(0)$ that, beginning with density $\rho(\bm{\eta}(0)) = x$ end up after $T$ time steps as a configuration $\bm{\eta}(T)$ with density $\rho(\bm{\eta}(T))$ equal to $0$, $1/2$ or $1$ depending on whether the initial density was $x<1/2$, $x=1/2$, or $x>1/2$. For $\Phi^{184}_{232}(\varepsilon', \varepsilon'')$, we fix $T = T_{184}+T_{232}$ and define the efficiencies as
\begin{equation}
\label{eq:eff}
E_{\Phi}(\varepsilon', \varepsilon'')(x) =
\left\langle
\frac{{\#}\big\{\bm{\eta}(T): \rho(\bm{\eta}(T)) = \theta(x-1/2)\big\}}
{{\#}\big\{\bm{\eta}(0): \rho(\bm{\eta}(0)) = x \big\}}
\right\rangle,
\end{equation}
where $\theta(\cdot)$ is the Heaviside step function and the angle brackets denote average over the configuration space as well as over realizations of the noise. Notice that the case of $L$ even and $x=1/2$ is automatically accounted for the efficiency of the CA or PCA, in agreement with the behavior of the noiseless $\Phi^{184}_{232}$. It is also useful to define the average efficiencies
\begin{subequations}
\begin{equation}
E_{\Phi}^{(0)}(\varepsilon', \varepsilon'') = 
\langle E_{\Phi}(\varepsilon', \varepsilon'')(x < 1/2) \rangle,
\end{equation}
\begin{equation}
E_{\Phi}^{(1)}(\varepsilon', \varepsilon'') = 
\langle E_{\Phi}(\varepsilon', \varepsilon'')(x > 1/2) \rangle,
\end{equation}
\begin{equation}
E_{\Phi}(\varepsilon', \varepsilon'') = 
\langle E_{\Phi}(\varepsilon', \varepsilon'')(x) \rangle,
\end{equation}
\end{subequations}
together with $E_{\Phi}^{(\frac{1}{2})}(\varepsilon', \varepsilon'') = \langle E_{\Phi}(\varepsilon', \varepsilon'')(x=1/2) \rangle$ when $L$ is even. Notice that the overall efficiency $E_{\Phi}(\varepsilon', \varepsilon'')$ is a simple average over the performance of the automaton at all densities probed. It cannot be compared with the efficiencies obtained using initial configurations sampled close to some specific density or distributed according to some specific probability distribution other than the flat one over $0 \leq \rho \leq 1$.

\begin{table}
\caption{\label{tab:pca}Rule tables for PCA $\Phi^{184}_{232}(\varepsilon', \varepsilon'')$. Reads like Table~\ref{tab:line}, except that the central bit of the initial neighborhood reaches its final state with the probability given at the leftmost column.}
\medskip
\begin{tabular}{ccccccccc}
\hline\hline
PCA~184:         & 111 & 110 & 101 & 100 & 011 & 010 & 001 & 000 \\ \hline
$1-\varepsilon'$ &  1  &  0  &  1  &  1  &  1  &  0  &  0  &  0  \\
$\varepsilon'  $ &  0  &  1  &  0  &  0  &  0  &  1  &  1  &  1  \\ \hline\hline\\
\hline\hline
PCA~232:          & 111 & 110 & 101 & 100 & 011 & 010 & 001 & 000 \\ \hline 
$1-\varepsilon''$ &  1  &  1  &  1  &  0  &  1  &  0  &  0  &  0  \\
$\varepsilon''  $ &  0  &  0  &  0  &  1  &  0  &  1  &  1  &  1  \\ \hline\hline
\end{tabular}
\end{table}

In this article we investigate the classification performance of $\Phi^{184}_{232}(\varepsilon', \varepsilon'')$ for one-dimensional arrays of cells under periodic boundary conditions. We analyze systems of several lengths, both odd and even, because contrary to the GKL and related CA, the behavior of $\Phi^{184}_{232}$ is well defined when $L$ is even and we want to see whether and by how much $\Phi^{184}_{232}(\varepsilon', \varepsilon'')$ inherits the performance of the noiseless version at $\rho=1/2$. In our simulations, we pick $1000$ sample configurations for each value of $\rho$ chosen and average the classification performance of $\Phi^{184}_{232}(\varepsilon', \varepsilon'')$ for each sampled configuration over $1000$ realizations of the noise. Noise is thus dynamic in our simulations, not quenched. A typical efficiency profile with fixed $L$, $\varepsilon'$, and $\varepsilon''$ is traced out of data for ${\sim}100$ different values of $\rho$. This amount of sampling suffices to provide smooth curves and useful figures. The uncertainties in the figures are mostly of the order of the sizes of the symbols used to display them. The reader may assume that the errors are typically of the order of $\pm 2\%$ of the main figure given. The results of our experiments are presented in the next section.


\section{\label{results}Results of the numerical experiments}

\subsection{\label{efficiency}The efficiency of the PCA}

\begin{figure}
\begin{tabular}{cc}
\includegraphics[viewport=71 88 488 762, scale=0.22, angle=-90]{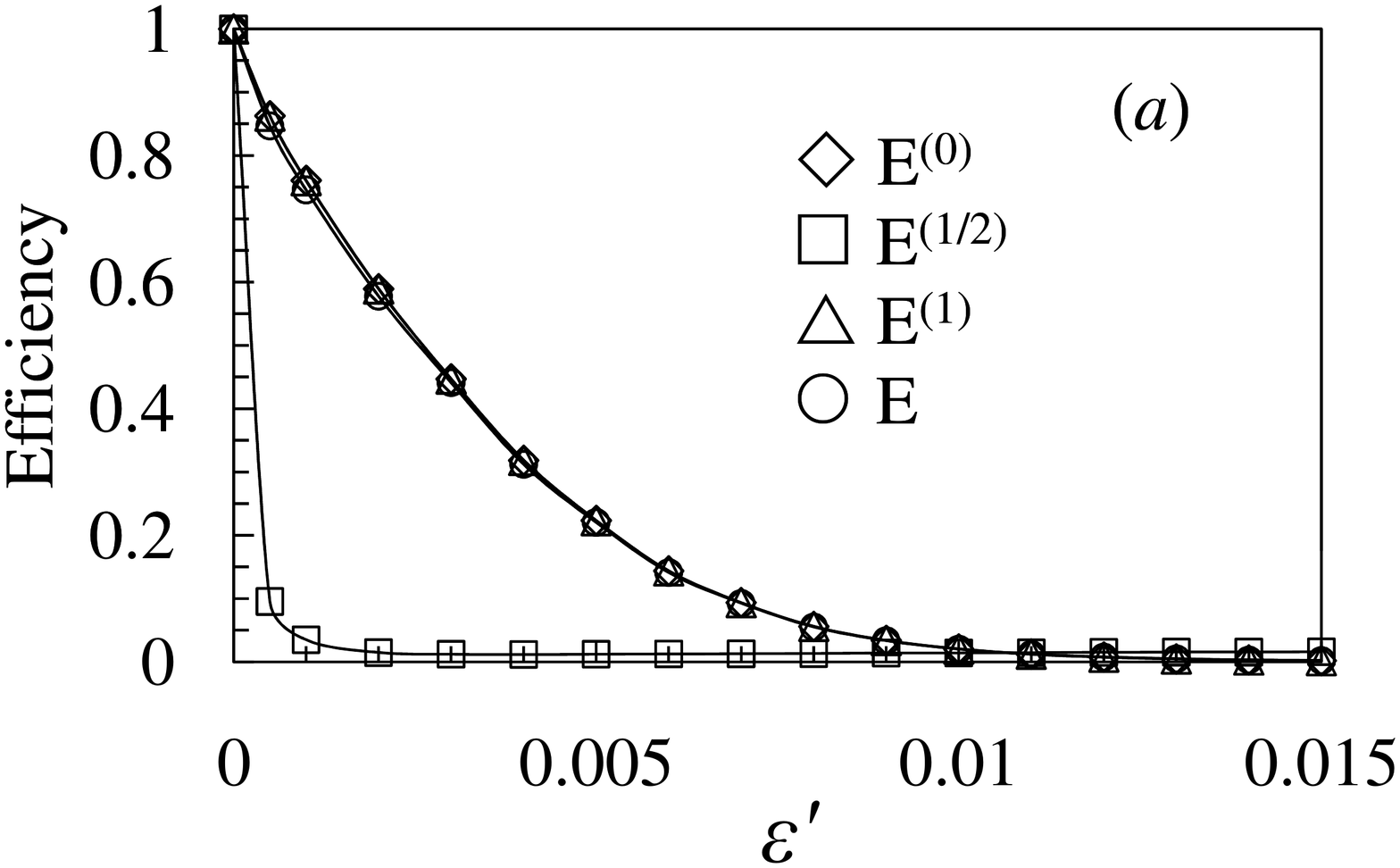} \\
\includegraphics[viewport=71 88 488 763, scale=0.22, angle=-90]{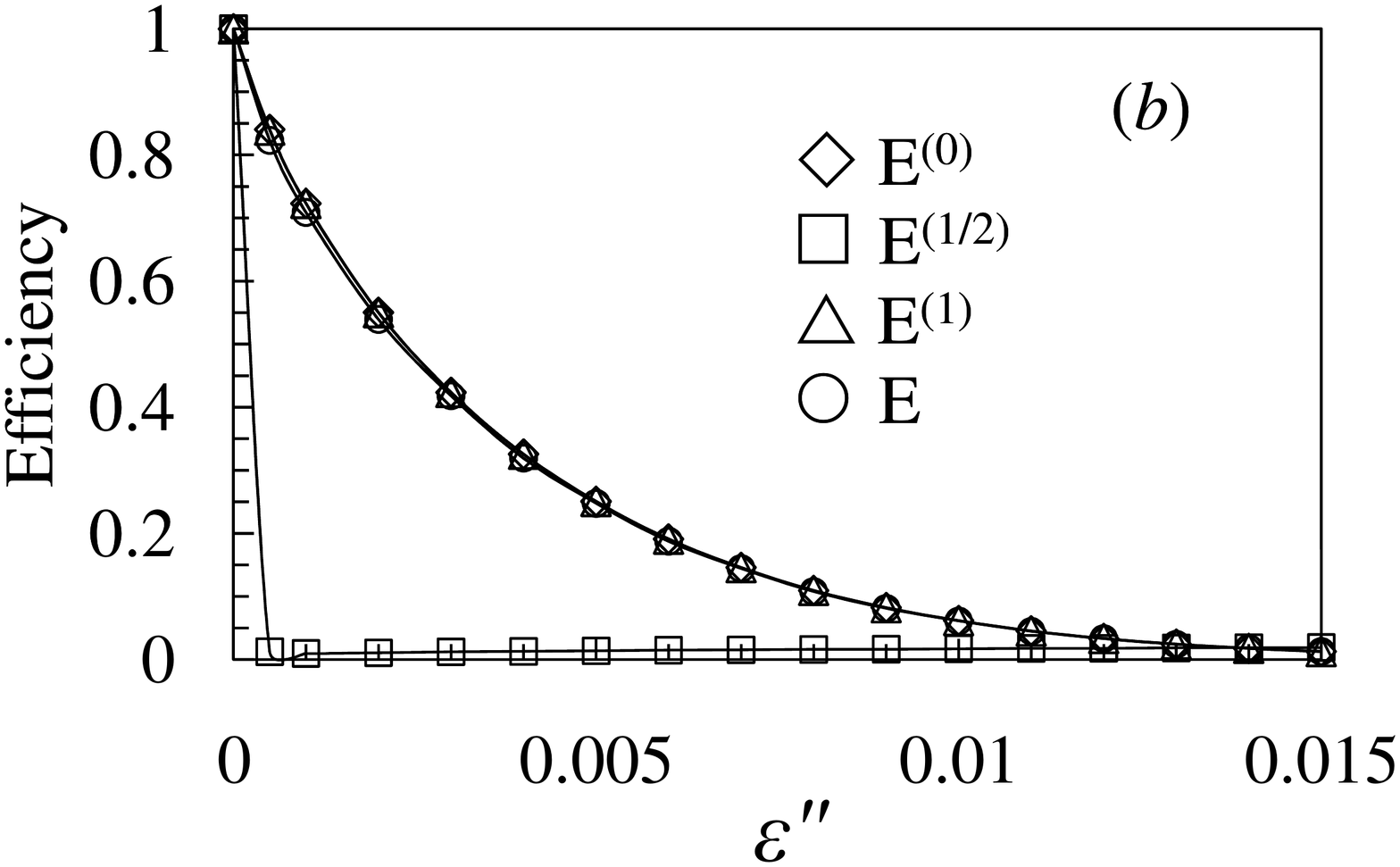}
\end{tabular}
\caption{\label{fig:eps}Efficiencies of $\Phi^{184}_{232}(\varepsilon', \varepsilon'')$ along the lines ($a$) $\varepsilon''=0$ and ($b$) $\varepsilon'=0$ in a PCA of $L=150$ cells. Notice how $E_{\Phi}^{(0)}$, $E_{\Phi}^{(1)}$, and $E_{\Phi}$ superimpose greatly and cannot be discerned clearly in the graphs. This results from the symmetry $E_{\Phi}^{(0)} \leftrightarrow E_{\Phi}^{(1)}$ about $\rho=1/2$ and the steep decay of $E_{\Phi}^{(\frac{1}{2})}$ toward zero with increasing noise.}
\end{figure}

\begin{figure}
\begin{tabular}{c}
\includegraphics[viewport=123 54 472 792, scale= 0.28, angle=-90]{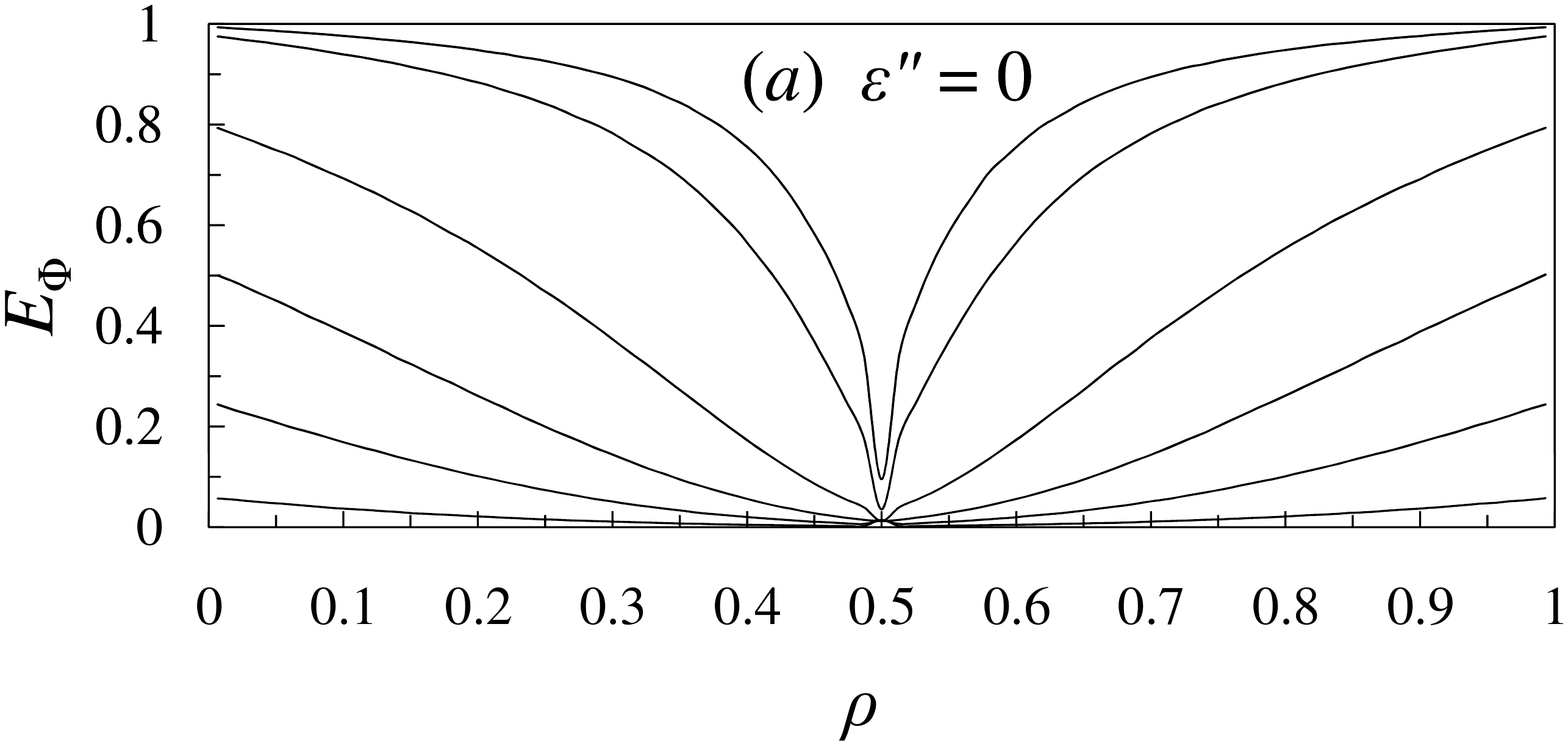} \\
\includegraphics[viewport=123 56 473 792, scale= 0.28, angle=-90]{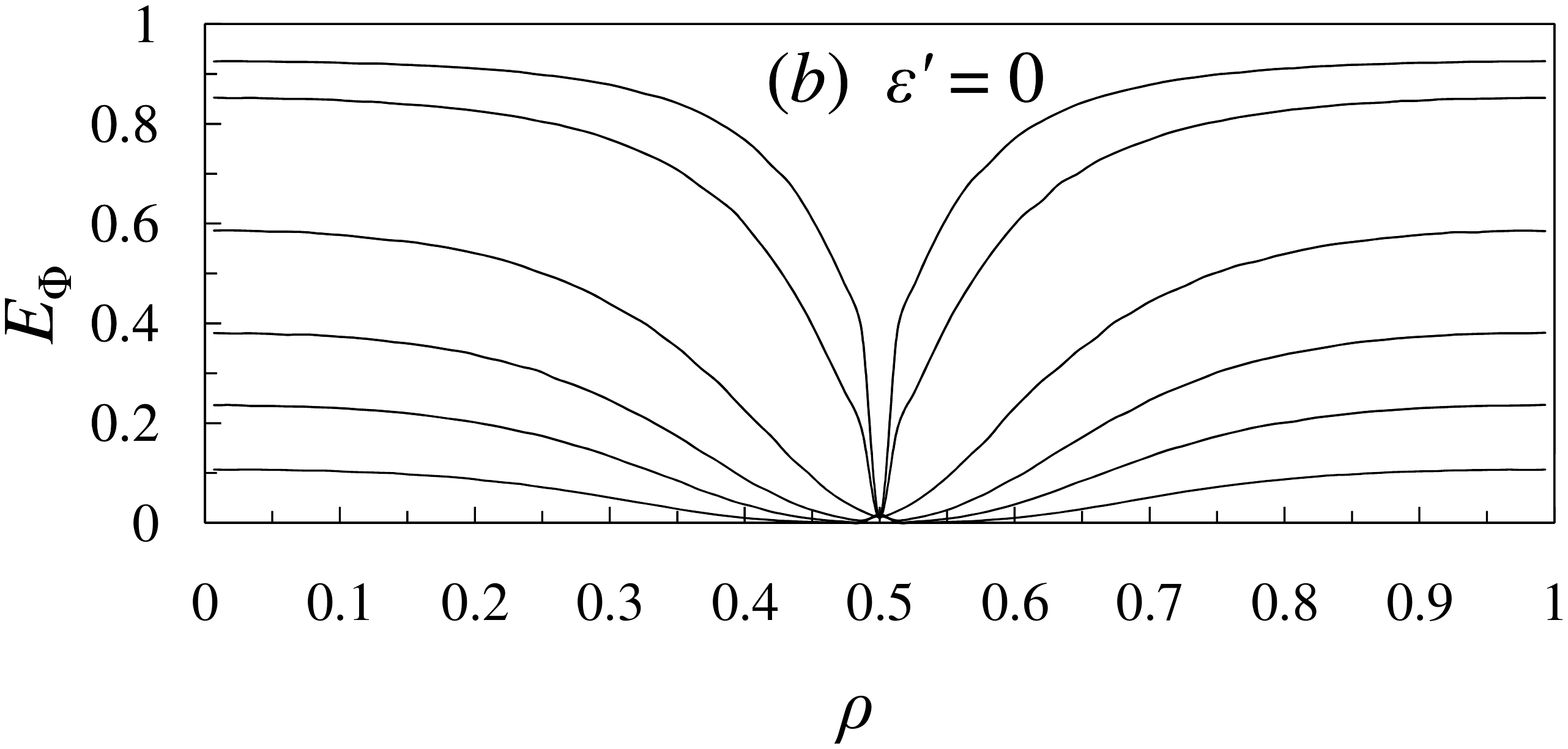} 
\end{tabular}
\caption{\label{fig:rho}Overall efficiencies $E_{\Phi}(\rho)$ for $\Phi^{184}_{232}(\varepsilon', \varepsilon'')$ along the lines ($a$) $\varepsilon''=0$ and ($b$) $\varepsilon'=0$ in a PCA of $L=150$ cells, as in Fig.~\ref{fig:eps}. From the uppermost curve downward, $\varepsilon'$ or $\varepsilon''$ equals $0.0005$, $0.001$, $0.003$, $0.005$, $0.007$, and $0.01$.}
\end{figure}

We begin by considering the case of an even number of cells, $L=150$, just to verify how $E_{\Phi}^{(\frac{1}{2})}$ behaves. The efficiencies of $\Phi^{184}_{232}(\varepsilon', \varepsilon'')$ in this case subject to noise in the interval $0.0005 \leq \varepsilon', \varepsilon'' \leq 0.015$ along the lines $\varepsilon''=0$ and $\varepsilon'=0$ appears in Fig.~\ref{fig:eps}. Total efficiency profiles for several different levels of noise appear in Fig.~\ref{fig:rho}.

\begin{figure}
\includegraphics[viewport=122 66 451 793, scale=0.32, angle=-90]{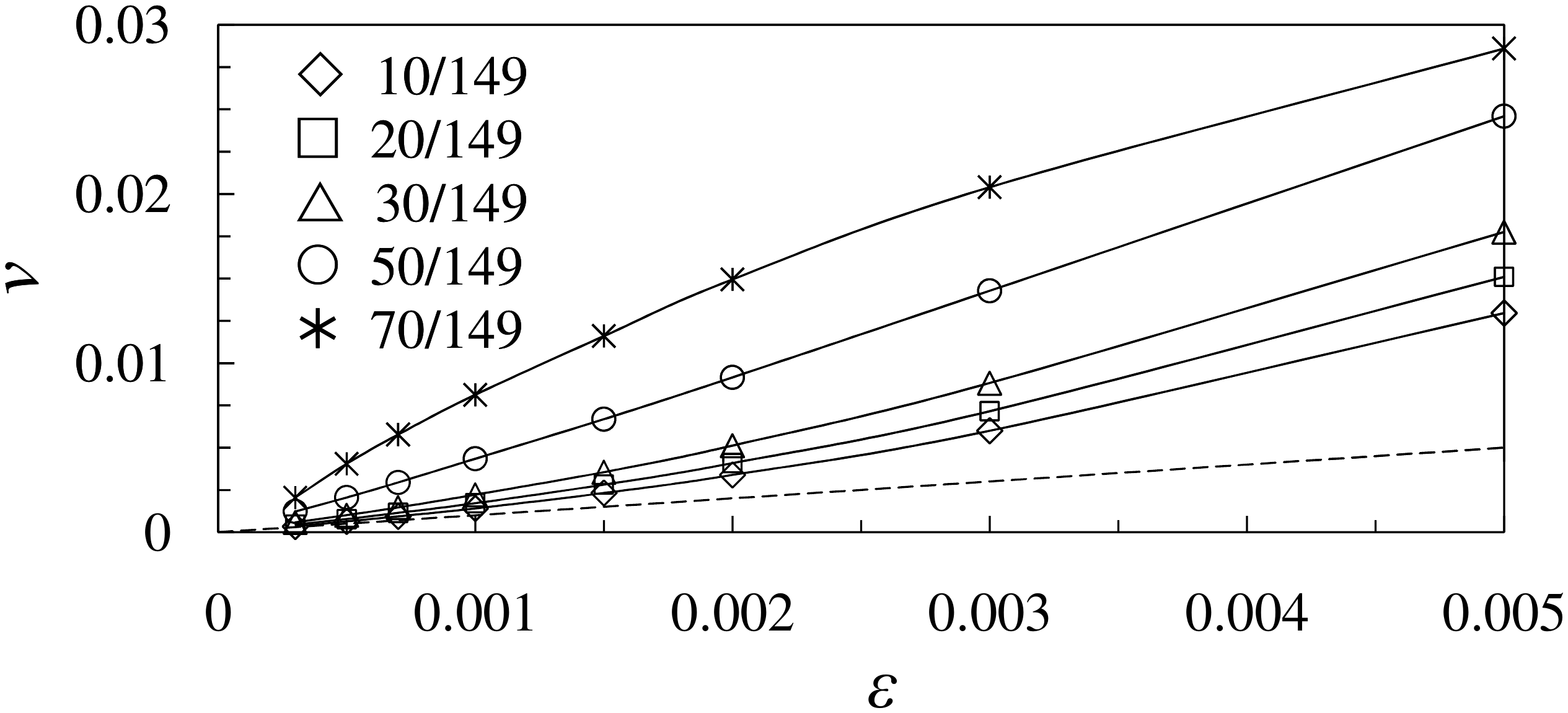}
\caption{\label{fig:loc}{\it Loci\/} of ``isoefficiency'' between $\Phi^{184}_{232}(\varepsilon)$ and $\Phi_{\rm GKL}(\nu)$, Eq.~(\ref{eq:loc}), for a few choices of $\rho$ (indicated as fractions). Notice the different ranges of the axes. For fixed $\rho$, $\Phi^{184}_{232}(\varepsilon)$ performs worse than $\Phi_{\rm GKL}(\nu)$ for any noise levels ($\varepsilon, \nu)$ beneath the respective isoefficiency curve. The line $\nu = \varepsilon$ (dashed) is shown for reference.}
\end{figure}

The data reveal that $\Phi^{184}_{232}(\varepsilon', \varepsilon'')$ is highly sensitive to noise. From Fig.~\ref{fig:eps}, it is clear that already at very small levels of noise its classification performance becomes significantly degraded and barely compares with the average performance of the noiseless GKL automaton (namely, ${\sim}81.6\%$ around $\rho=1/2$ and ${\sim}97.7\%$ overall), an imperfect classifier. This observation posed us to compute the solutions of the equation
\begin{equation}
\label{eq:loc}
E_{\Phi}(\varepsilon)(\rho) = E_{\rm GKL}(\nu)(\rho)
\end{equation}
along the line $\varepsilon' = \varepsilon''$, where $E_{\rm GKL}(\nu)(\rho)$ is the efficiency of the GKL automaton under noise $0 \leq \nu \leq 1/2$ (cf.~Appendix~\ref{app-b} for the definition of this automaton). In order to solve the above equation, we fix $\rho$ and $\varepsilon$ and find the $\nu$ that satisfies it. In this comparison test, we ran both automata with $L=149$ cells and allowed only $T=L$ time steps for the GKL automaton to classify the densities, when the customary is to run it through $T=2L$ time steps. In principle this grants some advantage to $\Phi^{184}_{232}(\varepsilon', \varepsilon'')$, but we found it negligible. A few {\it loci\/} of Eq.~(\ref{eq:loc}) for different choices of $\rho$ are shown in Fig.~\ref{fig:loc}. We invariably found $\nu > \varepsilon$, meaning that $\Phi_{\rm GKL}(\nu)$ is more robust and classifies density more reliably that $\Phi^{184}_{232}(\varepsilon', \varepsilon'')$ under the influence of noise, whether over the line $\varepsilon' = \varepsilon''$ or along nearby paths.

Noise is particularly deleterious to the performance of $\Phi^{184}_{232}(\varepsilon', \varepsilon'')$ close to $\rho=1/2$, a region where $\Phi_{\rm GKL}(\nu)$ proved far more robust. From Fig.~\ref{fig:loc} we see that along most of the curve for $\rho=70/149$, $\Phi_{\rm GKL}(\nu)$ can cope with as much as eight times more noise than $\Phi^{184}_{232}(\varepsilon', \varepsilon'')$ for the same performance, and the trend seems to indicate that this factor can become even greater at higher levels of noise or closer to $\rho=1/2$. Notice, however, that increasing levels of noise ``disrupt'' the PCA at some point, above which $E_{\Phi} \to 0$. In an automaton of $L=149$ cells, this point of rupture for $\Phi^{184}_{232}(\varepsilon', \varepsilon'')$ along the line $\varepsilon' = \varepsilon''$ is close to $\varepsilon \simeq 0.0152$ and for $\Phi_{\rm GKL}(\nu)$ it is close to $\nu \simeq 0.053$. Notice that the results reported in \cite{maes} regarding the ergodicity of the noisy GKL automaton were obtained mostly beyond its point of rupture.

Figures~\ref{fig:eps} and \ref{fig:rho} reveal that $\Phi^{184}_{232}(\varepsilon', \varepsilon'')$ is slightly more sensitive to noise in its CA~232 second half than in its CA~184 first half. Otherwise, for initial densities far off the critical region around $\rho=1/2$ the CA~232 part is more robust to noise, in the sense that when the noisy CA~$232$ part receives a configuration preprocessed by a noiseless CA~$184$ part it misses the correct classification less than when it receives a configuration preprocessed by a noisy CA~$184$ with the same level of noise and processes it without noise. 
The case most sensitive to noise is $\rho=1/2$, as it can be verified from the behavior of $E_{\Phi}^{(\frac{1}{2})}$. This is an expected treat, since initial states with $\rho=1/2$ are the ones for which the smallest error favoring either side of the density can most probably lead to an erroneous classification.

Put together, Figs.~\ref{fig:eps} and \ref{fig:rho} allow us to conclude that except for very small error rates, say, $\sqrt{\varepsilon' \varepsilon''} < 0.001$, and for input strings with a very definite density, say $|\rho - 1/2| > 1/3$, PCA $\Phi^{184}_{232}(\varepsilon', \varepsilon'')$ is an unreliable density classifier. Figure~\ref{fig:loc} also tells us that the GKL automaton is a better option for classifying density in noisy environments or using faulty components.

\subsection{\label{ergodic}The ergodicity of the PCA}

Clearly, $\Phi^{184}_{232}(\varepsilon'=0, \varepsilon''=0)$ is nonergodic. When $\varepsilon' \neq 0$ or $\varepsilon'' \neq 0$ or both, however, we can ask whether the PCA sweeps through the configuration space or gets clogged in some finite neighborhood, signaling nonergodicity. We then define $\tau_{\Phi}(L, \varepsilon', \varepsilon'')$ as the time it takes the initial configuration of all $1$'s (not absorbing if $\varepsilon'$ or $\varepsilon''$ or both are non-null) to evolve into a configuration with density $\rho < 1/2$. We take $\rho=1/2$ as the threshold because when the system reaches a configuration with this density it has, in an intuitive sense, ``crossed the barrier to the other side of the well.'' For $\Phi^{184}_{232}(\varepsilon', \varepsilon'')$, there is the problem of defining the shares of each CA of the assembly line in this relaxation time. Here we take $T_{184} = T_{232}$ over increasing time spans. Beginning with some small $T_{184} = T_{232}$, we evolve the initial configuration through $T = T_{184} + T_{232}$ time steps, measure the density and, if $\rho \geq 1/2$, increment $T_{184} \leftarrow T_{184}+1$, $T_{232} \leftarrow T_{232}+1$, run the PCA with an initial configuration of all $1$'s through the updated $T = T_{184} + T_{232}$ and measure $\rho$ again, and repeat this procedure until $\rho < 1/2$ for some $T$. We then estimate $\tau_{\Phi}(L, \varepsilon', \varepsilon'')$ from an average over $1000$ such hitting times.

We first estimated $\tau_{\Phi}(L, \varepsilon', \varepsilon'')$ along the line $\varepsilon' = \varepsilon''$. Following \cite{maes}, we look for flipping times of the form
\begin{equation}
\label{eq:tau}
\tau_{\Phi}(L,\varepsilon) \sim \exp(f(L,\varepsilon)),
\end{equation}
possibly with an algebraic prefactor. The putative relation (\ref{eq:tau}) is drawn on an analogy between the space-time diagram of the PCA and the configuration of a $2D$ interacting classical spin-$\frac{1}{2}$ model over $\Lambda = \{(l,t) \in \{1, 2, \ldots, L\} \times \mathbb{N}\}$. In this analog $2D$ spin model, close to the critical point $0 < |T-T_{c}|/T_{c} \ll 1$ the correlation length in the $t$ direction scales like $\xi_{\|}(L,T) \sim |T-T_{c}|^{-\nu_{\|}}$ and also like $\xi_{\|}(L,T) \sim \exp(L^{z}/T)$ for $T<T_{c}$ and $\xi_{\|}(L,T) \sim L^{z}$ at $T=T_{c}$, being bounded in $L$ for $T>T_{c}$ \cite{binney,marro}. The real numbers $\nu_{\|}$ and $z$ are critical exponents. The $2D$ spin model scenario is translated into the PCA scenario by identifying $\varepsilon$ with $T$ and $\tau_{\Phi}(L,\varepsilon)$ with $\xi_{\|}(L,T)$, with the ergodic phase of the PCA corresponding to the disordered ($T>T_{c}$) phase of the spin model. A nonergodic dynamics would thus imply that $\tau_{\Phi}(L, \varepsilon)$ diverges as $L\!\uparrow\!\infty$ with finite $\varepsilon$'s, while for an ergodic dynamics $f(L,\varepsilon)$ should remain bounded in $L$. Obviously, $\tau_{\Phi}(L,\varepsilon)$ is expected to diverge as $\varepsilon \!\downarrow\! 0$.

\begin{figure}
\includegraphics[viewport=132 57 454 793, scale=0.32, angle=-90]{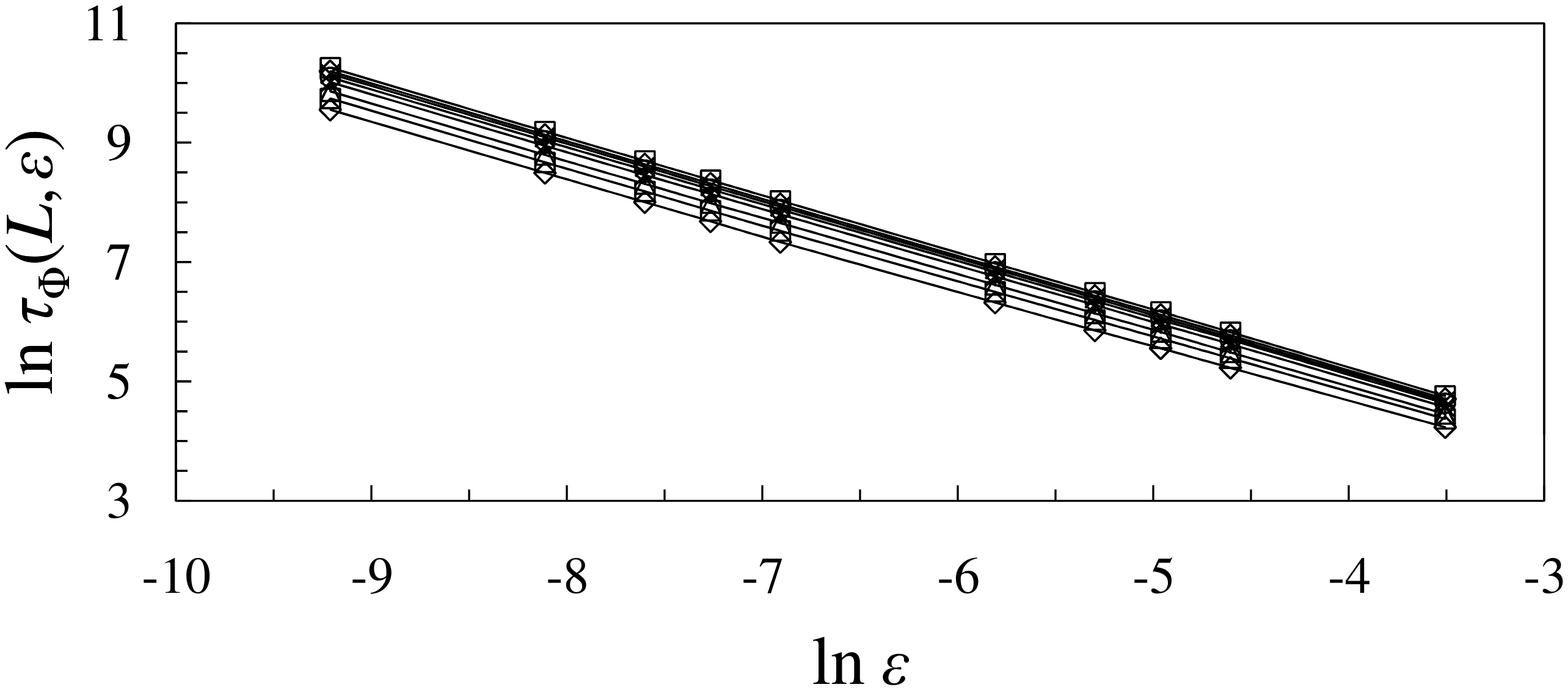}
\caption{\label{fig:tau-eps}Relaxation times $\tau_{\Phi}(L,\varepsilon)$ for some fixed $L$ between $L=149$ (lowermost curve) and $L=3499$ (uppermost curve); cf.~Fig.~\ref{fig:tau-l}.}
\end{figure}

\begin{figure}
\includegraphics[viewport=122 58 463 794, scale=0.32, angle=-90]{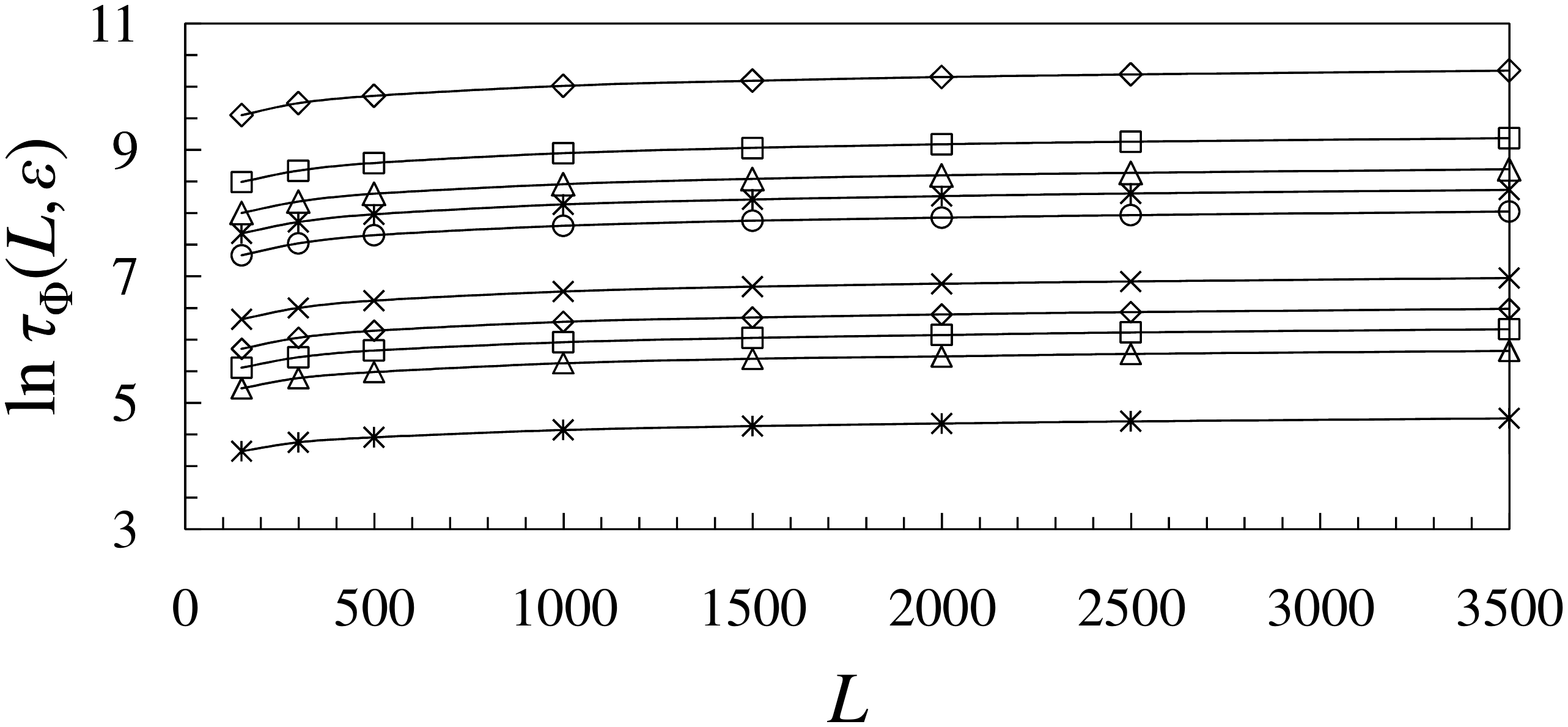}
\caption{\label{fig:tau-l}Relaxation times $\tau_{\Phi}(L, \varepsilon)$ for some fixed $\varepsilon$ between $\varepsilon=0.0001$ (uppermost curve) and $\varepsilon=0.03$ (lowermost curve).}
\end{figure}

Plots of $\ln \tau_{\Phi}(L, \varepsilon)$ for fixed odd values of $L$ and $\varepsilon$ appear in Figs.~\ref{fig:tau-eps} and \ref{fig:tau-l}. The errors in the data are in the range of $5$--$10$\%, with higher uncertainties for smaller $L$ and larger $\varepsilon$. We found that for fixed $L$, $\ln \tau_{\Phi}(L,\varepsilon)$ varies with $\ln \varepsilon$. Both the conditions that $\tau_{\Phi}(L,\varepsilon)$ diverges as $\varepsilon \!\downarrow\! 0$ and that $f(L,\varepsilon)$ remains bounded in $L$ are observed by the data. Moreover, from Fig.~\ref{fig:tau-eps} we already see that $f(L,\varepsilon)$ is quite insensitive to $L$---actually, the curves clump together more as $L$ gets larger, indicating sublinear growth of $f(L,\varepsilon)$ with $L$. Linear fits to the curves give slopes $0.91 \leq \alpha_{L} \leq 0.97$, all with correlation coefficients $R^{2} > 0.9998$. An extrapolation of the regression coefficients $\alpha_{L} \times L^{-1}$ gives $\alpha_{\infty} = 0.957 \pm 0.004$ with a correlation coefficient $R^{2} \simeq 0.89$. This provides considerable evidence for the ergodicity of $\Phi^{184}_{232}(\varepsilon', \varepsilon'')$, at least on the line $0.0001 \leq \varepsilon' = \varepsilon'' \leq 0.03$. Complementary evidence comes from Fig.~\ref{fig:tau-l}, where we see that, for fixed $\varepsilon$, the relaxation times are largely independent of $L$. The interpretation of this behavior is that no matter how far the initial configuration $11 \cdots 1$ is from the typical stationary configurations, the system is able to reach them in finite (and, actually, relatively short) times, forgetting its initial configuration. In all cases we found that $\tau_{\Phi}(L, \varepsilon', \varepsilon'') \sim \varepsilon^{-\alpha}$ (with $\varepsilon = \varepsilon'$ or $\varepsilon''$---it does not really matter) and exponent in the range $0.91 \leq \alpha \leq 0.98$.

Altogether, these facts indicate that $\Phi^{184}_{232}(\varepsilon', \varepsilon'')$ is probably ergodic for nonzero levels of noise.

We would like to remark that the exponents we found for $\tau_{\Phi}(L, \varepsilon', \varepsilon'')$ are all very close to the mean-field value $\nu_{\|} = 1$ expected for a completely uncorrelated behavior. The fact that the PCA is very sensitive to noise is an indication that its stationary states in the ergodic region are very uncorrelated in space and time, and we guess that its critical behavior in the disordered phase is probably ruled by mean-field exponents. We repute the deviation of the values of $\alpha$ observed in our simulations from the exact mean-field value not only to statistical errors but also to finite-size effects---indeed, we found larger $\alpha_{L}$ for larger $L$ and $\varepsilon$.

\section{\label{summary}Summary and further developments}

We conclude that while $\Phi^{184}_{232}$ can classify density perfectly, its noisy version is highly sensitive to noise, almost to the point of being useless. Since error detection and correction are not available to locally interacting CA or PCA, any real, noisy-world application of $\Phi^{184}_{232}$ has to ponder its sensitivity to noise and the ensuing limitations. We have also accumulated evidence for the ergodicity of $\Phi^{184}_{232}(\varepsilon', \varepsilon'')$ at any finite level of noise, meaning that the mechanism that drives the CA into its absorbing configurations does not survive to noise. The critical behavior of the PCA close to and above the critical point $(\varepsilon', \varepsilon'')_{c} = (0,0)$ is probably ruled by mean-field exponents.

It would be interesting to investigate the robustness of $\Phi^{184}_{232}(\varepsilon', \varepsilon'')$ in more general graphs. Preliminary results indicate that single CA rules do not perform well under such conditions \cite{tomassini,darabos}, but these results do not extend {\it prima facie\/} to composite CA like $\Phi^{184}_{232}$. A study of the efficiency of the GKL rule under the influence of both noise and graph topology was carried out in \cite{amaral}, with the conclusion that the noisy GKL rule performs worse than a simple majority rule in small-world networks. Results obtained for random Boolean networks also point to a better performance for these models than for CA or PCA \cite{boolean}. Notice that while the CA~232 part of $\Phi^{184}_{232}$ is the majority rule that can be easily extended to any node topology, the CA~184 part would have to be redesigned to work with nodes of degree $k>2$. Analytical approaches to $\Phi^{184}_{232}(\varepsilon', \varepsilon'')$ are also desirable, although rigorous results for PCA are hard to obtain. Mean-field approximations, however, are amenable to analysis and may help to set forth new results for this class of PCA \cite{bollobas}.


\begin{acknowledgments}

The author is grateful to Prof.~M\'{a}rio J. de Oliveira (IF/USP) for having pointed an embarassing mistake in a previous version of the manuscript and for continuous encouragement.

\end{acknowledgments}


\appendix*

\section{\label{app-b}The noisy GKL automaton}

In Sec.~\ref{efficiency} we were led to investigate the noisy GKL automaton.  For the sake of completeness, we specify it here.

The noiseless GKL automaton evolves according to the following rules \cite{gkl}: if $\eta_{i}(t)=0$, then $\eta_{i}(t+1) = \Phi_{\rm GKL}(\eta_{i}(t))  = \theta(\eta_{i-3}(t)+\eta_{i-1}(t)+\eta_{i}(t)-3/2)$, where $\theta(\cdot)$ is the Heaviside step function introduced in Eq.~(\ref{eq:eff}), and if $\eta_{i}(t)=1$, then $\eta_{i}(t+1) = \Phi_{\rm GKL}(\eta_{i}(t)) = \theta(\eta_{i}(t)+\eta_{i+1}(t)+\eta_{i+3}(t)-3/2)$---i.e., $\eta_{i}(t+1)$ equals the majority of its particular neighborhood at instant $t$, that itself depends whether $\eta_{i}(t)=0$ or $1$. The noisy version just adds the possibility that, at every site, instead of $\eta_{i}(t+1) = \Phi_{\rm GKL}(\eta_{i}(t))$, with probability $\nu$ we have $\eta_{i}(t+1) = 1-\Phi_{\rm GKL}(\eta_{i}(t))$ \cite{maes}.


\end{document}